\documentclass[a4paper,11pt]{article}
\pdfoutput=1 
\usepackage{jcappub}

\usepackage[T1]{fontenc} 
\usepackage{mathrsfs}
\usepackage{color}
\usepackage{slashed}
\usepackage{epsfig}
\usepackage{enumerate}
\usepackage{color}
\usepackage{wrapfig}
\usepackage[caption=false]{subfig}
\usepackage{breqn}

\title{\boldmath  Anomalous X-Ray Galactic signal from $7.1$ keV spin-$3/2$ dark matter decay}
\author[a,\,\dagger]{Sukanta Dutta,}
\author[b,\,\,\star]{Ashok Goyal,}
\author[b,\,\,\dagger\dagger]{Sanjeev Kumar}
\affiliation[a]{SGTB Khalsa College, University of Delhi, Delhi, India.}
\affiliation[b]{Department of Physics $\&$ Astrophysics, University of Delhi, Delhi, India.}
\emailAdd{$\dagger$sukanta.dutta@gmail.com}\emailAdd{$\star$agoyal45@yahoo.com}\emailAdd{$\dagger\dagger $sanjeev3kumar@gmail.com}

\abstract{
In order to explain the recently reported peak at $3.55$ keV in the galactic
x-ray spectrum, we propose a simple model. In this model, the Standard Model
is extended by including a neutral spin-$3/2$ vector-like fermion
that transforms like a singlet under SM gauge group.
This $7.1$ keV spin-$3/2$ fermion is considered to comprise a portion of the observed dark matter. Its decay into a neutrino and a photon with decay life commensurate 
with the observed data, fits the relic dark matter density and obeys the astrophysical
constraints from the supernova cooling.
}
\begin{document} 
\maketitle
\flushbottom

\section{Introduction}

Recently X-Ray emission at $\sim 3.55$ keV  has been observed in the XMM- Newton X-Ray observatory \cite{Bulbul:2014sua,Boyarsky:2014jta} in many Galaxy clusters and in the Andromeda Galaxy spectra.  The observed flux and any X-Ray line energy measured in the MOS spectra is given by 
\begin{eqnarray}
\Phi_\gamma^{\rm MOS}&=&4.0^{+0.8}_{-0.8} \times 10^{-6}\,\,  {\rm photons}\,\,{\rm cm}^{-2} \,\,{\rm sec}^{-1}\nonumber\\
E_\gamma ^{\rm MOS}&=&3.54 \pm 0.02\,\, {\rm keV}
\end{eqnarray}
The source of this line is yet to be identified. An attractive possibility, considered in the literature to explain the observed flux and energy, is to attribute it to the decay/ annihilation of some dark matter particle which is stable over cosmological time scale and can account for at least a significant fraction of dark matter relic density with mass and decay life time commensurate with the observed data. Sterile Neutrino of mass $7.1$ keV capable of producing warm Dark Matter (WDM) density through resonant or non-resonant production with parameters required to produce the observed signal is an attractive proposition discussed in the literature \cite{Ishida:2014dlp,Chakraborty:2014tma,Modak:2014vva,Abazajian:2014gza,Biswas:2015sva}. R-parity violating decays of the lightest super-symmetric  particle (LSP), the decay of gravitons and axions,
into neutrino photon pair and the decay of scaler field $\phi$ or axion-like pseudoscalar fields $ a$  into photon pairs as possible explanation of the signals have been considered in the literature \cite{Biswas:2015sva,Liew:2014gia,Kolda:2014ppa,Higaki:2014zua,Kong:2014gea,Covi:2009pq,Lee:2014xua, Krall:2014dba,Choi:2014tva,Babu:2014pxa} with varying success.  The scalar case is of particular interest because the scale of the new 
physics may involve super-symmetry (SUSY) which conforms to the expectations from the 
physics of moduli.

\par  Several  new physics models beyond the standard model 
(SM) predict the existence of spin-$3/2$ particles. In models of super-gravity, 
the graviton is accompanied by spin-$3/2$ gravitino super partner. In models of 
composites \cite{Burges:1983zg}, the top quark has an associated spin-$3/2$  resonance. 
New physics models may include exotic fermions and gauge bosons which are not 
present in the SM. Spin-$3/2$ fermions also exist as Kaluza -Klien modes in 
string theory \cite{ArkaniHamed:1998rs,Randall:1999ee} 
if one or more of  compactification radii are of the scale lower 
than the Planck scale. 
\par The prediction of spin-$3/2$ particle as a cold dark  matter has been made by several authors in   SUSY models \cite{Lemoine:2005hu, Jedamzik:2005ir}. Gravitinos with mass in the keV range have been studied as the probable WDM candidate in various SUSY models \cite{Baltz:2001rq,Fujii:2002fv,Moultaka:2006su,Gorbunov:2008ui} even before the observation of $3.55$ keV X ray emission. Recently, authors of the reference \cite{Ding:2012sm,Ding:2013nvx} have studied the implication of the effective four fermion interactions involving the DM spin-$3/2$ particle on relic
density, the antiproton to proton flux ratio in cosmic rays, and the elastic scattering off nuclei ( direct detection)  in the effective field theory approach. Constraints from direct detection of dark matter exist in literature on spin-$3/2$ WIMP candidates \cite{Savvidy:2012qa}. 

\par A recent comprehensive analysis by  the authors of   reference \cite{Jeltema:2014qfa} demonstrated  that
the measured flux of the $3.55$ keV line can be accounted for, by the conventionally known plasma lines without invoking the dark matter decay as its origin. This explanation, however, requires the fixing
of the abundances of different elements which are still uncertain to a certain degree.
We, thus, feel that it is worthwhile to investigate alternative interpretations that are consistent with the other astrophysical and cosmological data.

 \par In this paper, we consider a new neutral spin-$3/2$ fermion assumed to be
a vector-like SM singlet.  We will consider the decay of this $7.1$ keV DM particles into a neutrino-photon pair ($\chi \to \nu \gamma$) with
decay life commensurate with the observed galactic X-ray spectrum. This spin-
$3/2$  particle could exist as fundamental particle or could be a bound state 
of SM neutrino and $U(1)$ gauge bosons. We will explore the possibility of
such an exotic spin-$3/2$ particle to constitute the relic dark matter
for a reasonable choice of parameters and confront the model from cosmological
and astrophysical constraints.

\par In section \ref{spin3by2model}, we describe the spin-$3/2$ fermion  model. In section \ref{3.54keVline}, we discuss the implication of the model to explain the observed galactic X Ray spectrum data. In section \ref{relicspin3by2},
we obtain the relic abundance and the resulting constraints on the model
parameters. In section \ref{stellarloss}, we discuss the bounds obtained from 
supernova energy loss. Section \ref{results}  is devoted to results and 
discussion.
\section{The spin-$3/2$ Model}
\label{spin3by2model}
The standard model is extended by including a spin-$3/2$, vector like particle 
$\chi $, whose right-handed (RH) as well as left-handed (LH) projections transform the same way under $SU(2)\times SU(1)$. We further let $\chi$ to be a SM singlet. Spin-$3/2$ free Lagrangian is given by
\begin{eqnarray}
{\cal L}&=& \overline \chi_\mu (p_{\overline\chi})\,\, \Lambda^{\mu\nu} \,\,\chi_\nu (p_\chi) \,\,\,\,\,{\rm where}\nonumber\\
&&\Lambda^{\mu\nu}=\left(i\not\!\partial -m_\chi\right) g^{\mu\nu} - i \left( \gamma^\mu\gamma^\nu+\gamma^\nu\gamma^\mu\right) + i \gamma^\mu\not\!\partial \gamma^\nu + m_\chi\gamma^\mu\gamma^\nu.
\end{eqnarray}
Here $\chi_\mu $ satisfies $\Lambda^{\mu\nu} \chi_\nu =0$.  For on mass-shell $\chi$, we have
\begin{eqnarray}
 \gamma^\mu\chi_\mu(p_\chi)=0=\partial^\mu\chi_\mu(p_\chi)=(\not \!\! p-m_\chi) \chi_\mu(p_\chi).
\label{onmassshellcond}
\end{eqnarray}
The  spin-sum for spin-$3/2$ fermions 
\begin{equation}
{\cal S^+}_{\mu\nu}(p)=\sum_{i=-3/2}^{3/2} u^i_\mu (p)\,\,\overline u^i_\nu (p) 
 ~\text{and}~ 
{\cal S^-}_{\mu\nu}(p)=\sum_{i=-3/2}^{3/2} v^i_\mu (p)\,\,\overline v^i_\nu (p)
\end{equation}
are given by
\begin{eqnarray}
{\cal S}_{\mu\nu}^{\pm}(p)=-\,(\not\!\! p \pm m_\chi) \left[ g_{\mu\nu} -\frac{1}{3}\gamma_\mu\,\gamma_\nu-\frac{2}{3\,m_\chi^2} p_\mu \, p_\nu\mp\frac{1}{3m_\chi} \left(\gamma_\mu\, p_\nu-\gamma_\nu\, p_\mu\right)\right],
\end{eqnarray} 
respectively.

\par The most general leading order standard model gauge invariant interaction between the spin-$3/2$ SM singlet $\chi$ and SM spin-$1/2$ fermions is given by the effective dimension six operators:
\begin{eqnarray}
{\cal L}^{\rm Eff.}_{\rm Int.}&=&\sum_{i=1}^3\frac{C_i}{\Lambda^2}\,{\cal O}_i\nonumber\\
&=&\frac{C_1}{\Lambda^2} \,\,\overline l_L^k\,\gamma^\alpha \left[\gamma^\mu,\,\gamma^\nu\right]\, \chi_\alpha \tilde\phi \,B_{\mu\nu} + \frac{C_2}{\Lambda^2} \,\,\overline l_L^k\,\gamma^\alpha \,D_\mu \, \chi_\alpha D^\mu\tilde\phi + \frac{C_3}{\Lambda^2} \,\,\overline l_L^k \,\gamma^\mu\,\, \chi^\nu \tilde\phi \,B_{\mu\nu},
\end{eqnarray}
where $D_\mu\equiv i\, \partial_\mu-i\,(g_s/2)\, \lambda^a G^a_\mu-i\, (g/2)\, \tau^I\,W^I_\mu-i\, g^\prime \, Y\,B_\mu$,
$B_{\mu\nu}=(\cos\theta F_{\mu \nu}-\sin\theta Z_{\mu\nu})$, 
$\tilde\phi = i \tau_2 \phi$ and $l_L^k$ is the SM lepton doublet.
The weak  $U(1)$ hyper-charge $Y$ for  $\phi$ and $\chi$ are $1/2$ and $0$, respectively.

\par In view of the on-mass shell conditions as given in Eq. \eqref{onmassshellcond}, the second operator ${\cal O}_2$ vanishes and the third operator ${\cal O}_3$ becomes identical to the first operator ${\cal O}_1$. Therefore we are left with only one coupling constant $C$, which can be simplified to give after symmetry breaking:
\begin{eqnarray}
{\cal L}_I= \frac{C v_0}{\Lambda^2}  {\overline{\nu}_k}_L\left(p_{\nu_e}\right)\gamma_\mu \chi_\rho \left(p_\chi\right)\,\, \left(\cos\theta \,F^{\mu\rho}-\sin\theta\, Z^{\mu\rho}\right).\label{interaction}
\end{eqnarray}
Here $v_0$ is the SM Higgs vacuum expectation value and $\Lambda$ is the new cut-off scale.

\section{Galactic X-ray spectrum}
\label{3.54keVline}
The decay width for $\chi\to \nu_e\,\gamma$ is given by
\begin{eqnarray}
\Gamma_{\chi\to\nu_e\gamma}&=& \left(\frac{C v_0\,\cos\theta}{\Lambda^2}\right)^2 \,\frac{1}{16\pi}\,m_\chi^{3}\nonumber\\
&=&4.73\times 10^{-27} \,\,\left[\frac{C}{10^{-9}}\right]^2\,\left[\frac{\Lambda} {100\, {\rm TeV}}\right]^{-4}\,\,\left[\frac{m_\chi}{7\,{\rm keV}}\right]^3\,\,\,\,  {\rm sec}^{-1}.
\end{eqnarray}
Here we have taken the coupling of spin-$3/2$  particle  with only one generation (say for the first generation only) of SM neutrino.

\par The expected X-ray flux is proportional to the density of the decaying dark matter $\chi$. The WDM  which in the case considered here constitutes of spin-$3/2$ SM singlet, is believed to comprise a portion of the observed DM relic abundance with CDM as the dominant component \cite{Maccio':2012uh,Anderhalden:2012qt}. If the  X-ray galactic signal is interpreted as coming from the spin-$3/2$ WDM $\chi$ decaying into a neutrino and a photon pair, the required value of the life time of $\chi$ should be given by $\tau_\chi\sim  1.4\, f\times 10^{28}$ seconds, where $f$ ($0< f\le 1$) is the fraction of the relic dark matter density contributed by the WDM $\chi$. At $f=1.0$ the WDM $\chi$ would account for the entire dark matter relic density with a choice of new physics scale $\Lambda $ of the order of $\simeq 100$ TeV along with the coupling constant  $C\simeq 10^{-9}$. The small value of $C$ should not be
surprising as it can be considered to be a measure of trilinear
lepton number violating coupling and hence naturally small. Similar situation occurs in super-symmetric models of R-parity violating interactions considered in the literature  \cite{Kolda:2014ppa,Higaki:2014zua,Kong:2014gea,Covi:2009pq,Lee:2014xua,Krall:2014dba,Choi:2014tva,Babu:2014pxa} as possible explanation of the observed galactic X-ray flux. In realistic model, the mixing between photino and neutrino for example, is suppressed by a small parameter $\sim 10^{-10}$ characterising lepton number violation \cite{Krall:2014dba,Abazajian:2001vt,Lola:2007rw,Bomark:2014yja}.

\par If, the $7.1$ keV signal, on the other hand, is interpreted as coming from 
pair annihilation of $3.55$ keV spin-$3/2$ DM  into two photons, the annihilation cross-section $\langle\sigma\,v\rangle_{\rm ann.}$ has to match with the best-fit decay-width of $7.1$ keV  DM {\it i.e.} 
\begin{eqnarray}
\langle\sigma\,\,v\rangle_{\rm ann.}&\approx& 2\frac{ \Gamma_{\chi\to \nu_e\gamma}}{n_\chi},\,\,\,
\end{eqnarray}
where  $n_\chi= \rho_\chi / m_\chi\approx \left(10^4\,-\,10^5\right)$ cm$^{-3}$ is the number density of spin-$3/2$ DM. This translates into $\langle\sigma\,v\rangle_{\rm ann. fit}\simeq 2\times 10^{-16} \,\,\, {\rm GeV}^{-2}$. 
\par The spin-$3/2$ particles can couple to two photons through $U(1)$ gauge invariant dimension seven effective Lagrangian  
\begin{eqnarray}
{\cal L}_{\rm int.}= \frac{C_{\gamma\gamma}}{\Lambda^3}\,\overline\chi_\mu\,g^{\mu\nu}\,\chi_\nu\,\,\,F^{\alpha\,\beta}\,F_{\alpha\,\beta}.
\end{eqnarray}
This gives an annihilation cross-section 
$\sigma(\chi\overline\chi\to \gamma\gamma) \approx C^2_{\gamma\gamma}\,m_{\chi}^4/(\pi \Lambda^6)$ and the desired annihilation rate 
is achieved for $\Lambda \lesssim O(100)$ MeV (for $C_{\gamma\gamma} \lesssim 1$), which is
clearly unphysical. Thus, it is unlikely that the observed 
galactic X-ray signal can be explained by DM $\chi$'s
annihilation into photons. 
\section{Relic Abundance of Spin-$3/2$}
\label{relicspin3by2}
Since the $\chi$'s couple weakly to the SM particles and 
are nearly stable with a lifetime comparable to the age of the Universe if they  have to account for the observed X-ray flux,
they will decouple early when they are relativistic. They will,
therefore, contribute to the present mass density of the 
Universe as DM. Their abundance at decoupling is nearly
equal to the photon density at that time. During the adiabatic 
expansion of the Universe, their number densities remain 
comparable. A rough estimate of the bound on $\chi$ mass can be obtained 
just like the bound on the neutrino mass \cite{Cowsik:1972gh} by requiring
that the ratio of DM $\chi$ density to the critical density 
remain less than one. This gives $ m_{\chi} \le 12.8 g^\star \left(T_D\right)/g_{eff}$ eV. The effective number of degrees of freedom $g^\star\left(T_D\right)$ at decoupling  time of electroweak symmetry breaking transition  is found to be about  113.75. In the computation of 
$g^\star \left(T_D\right)$, we have included 
the effective degrees of freedom from all  SM particles and 
$\chi$, $\overline\chi$ spin-$3/2$ DM particles. However, in the
MSSM, $g^\star \left(T_D\right)$ is much larger $\sim 228.75$ and  thus it is  not reasonable for the spin-$3/2$ DM particle $\chi$ to have a mass of the order of about $7.1$ keV. 
\begin{wrapfigure}{l}{0.5\textwidth}
  \centering
\includegraphics[width=0.5\textwidth]{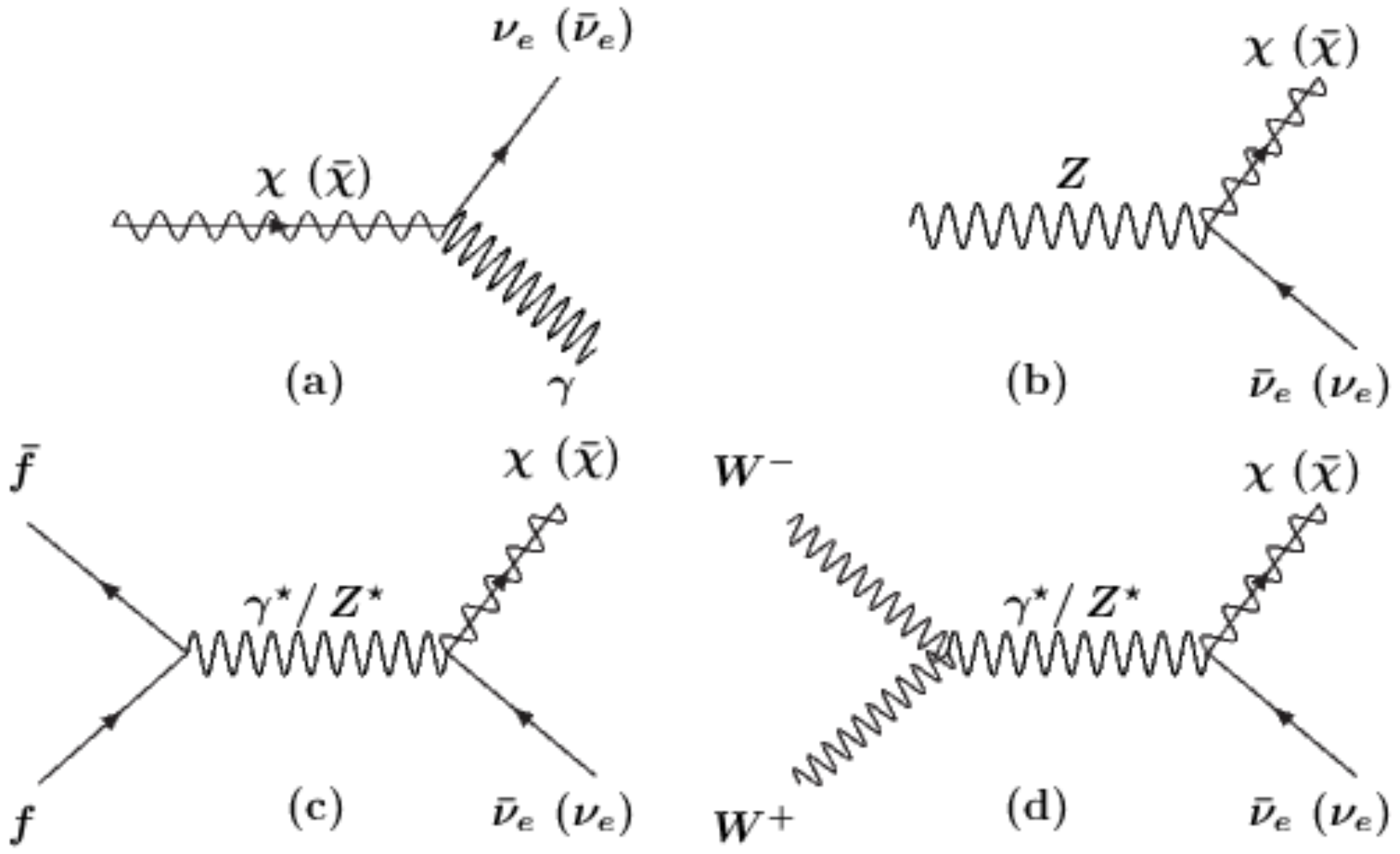}
\caption{\small \em{ Relevant diagrams for    decay and production of the DM candidate $\chi $. }}
\label{fig:feyndiags}
\end{wrapfigure}
The relic abundance of dark matter $\chi$ depends on
the sources of production of $\chi$ in the early Universe.
The leading order processes (shown in Figure \ref{fig:feyndiags}) that maintain
the DM $\chi$ in equilibrium with the rest of the SM plasma
are the decay rate of $Z \to \chi \overline\nu_e+ \overline\chi \nu_e$ and 
the $2 \to 2$ pair annihilation rates, namely,
$ \Gamma(Z \to \chi \overline\nu_e + \overline\chi \nu_e) $, $\Sigma_{f_i} \sigma (f_i \overline f_i \to \chi \overline\nu_e + \overline\chi \nu_e )$ and $\sigma (W^+ W^- \to \chi \overline\nu_e + \overline\chi \nu_e )$
where $\Sigma_{f_i}$  means summation over all
SM fermions (quarks and leptons). Using the interaction
Lagrangian given in Eq. \eqref{interaction}, the decay and
spin averaged annihilation cross-sections can be computed in a straightforward 
manner. We obtain:
\begin{equation}
\Gamma(Z \to \chi \overline\nu_e + \overline\chi \nu_e)
\approx \frac{C^2 v_0^2 \sin^2\theta}{\Lambda^4}
\frac{1}{72\pi}\frac{m_Z^2}{m_{\chi}^2}m_Z^3,\label{ZDecay}
\end{equation}

\begin{eqnarray}
&&\sigma \left(\sum_i f_i \overline f_i  \to \chi \overline\nu_e + \overline\chi \nu_e\right)
\approx  \frac{4 \pi \alpha C^2 v^2_0}{\Lambda^4} \frac{1}{128 \pi} 
\frac{8}{9}  \sum_i\frac{1}{\sqrt{1-\frac{4 m^2_i}{s}}}
\left(\frac{s}{m_\chi^2}\right)\, s^2 \times\nonumber\\ 
&&\hskip 2 cm \left[ \cos^2\theta \frac{{Q_f}_i^2}{s^2} \left( 1+
2\frac{m_i^2}{s}\right)+\frac{1}{\cos^2 \theta} \frac{1}{(s-m_Z^2)^2+\Gamma^2 m_z^2} 
\left\{ ({g_V^i}^2+{g_A^i}^2) \left( 1+\frac{3}{4}
\frac{s}{m_Z^2}\right)\right.\right.\nonumber\\
&&\hskip 2 cm + \left. \left. 2 \frac{m_i^2}{s} 
\left( ({g_V^i}^2-{g_A^i}^2)+\left\{\frac{3}{4} \left(\frac{s}{m_Z^2}\right)^2-\frac{3}{8}\frac{s}{m_Z^2}\right\}
 ({g_V^i}^2+{g_A^i}^2) \right) \right\} +\frac{{Q_f}_i (s-m_Z^2)}{s \lbrace (s-m_Z^2)^2
 +\Gamma^2 m_Z^2\rbrace }\right.\nonumber
\\&&\hskip 2 cm \left. \,\times \left\{ 
 g_v^i \left(\frac{3}{8}+\frac{m_Z^2}{s}+2 \frac{m_Z^2 m^2_i}{s^2} \right)
 +\frac{3}{8} \left(g^i_V+g^i_A \right) \frac{m^2_i}{s}  \left( \frac{2 m^2_i}{s}-1 \right) \right\} \right],
 \label{fermionAnnihilation}
\end{eqnarray}
and
\begin{eqnarray}
\sigma (W^+ W^- \to \chi \overline\nu_e + \overline\chi \nu_e)
\approx && \frac{4 \pi \alpha C^2 v^2_0}{\Lambda^4} \frac{1}{288 \pi} \left(1-\frac{4 m^2_f}{s}\right)^{-\frac{1}{2}}
\frac{1}{18} \frac{s}{m^2_\chi}\left(\frac{s}{m_W^2}\right)^2 s^2 \nonumber\\
&&\left[\frac{m_Z^2}{s} \frac{1}{(s-m_Z^2)^2+\Gamma^2 m_Z^2} 
\left( 1-\frac{4 m_W^2}{s}\right) 
 \left(1+ 20 \frac{m_W^2}{s} 
+12 \frac{m_W^4}{s^2}\right)\right.\nonumber\\
&&+\left. \frac{1}{s^2} \left( 1+16 \frac{m_{W}^2}{s} -68 \frac{m_W^4}{s^4}
-48 \frac{m_W^6}{s^3}\right) \right],\label{vectorFusion}
\end{eqnarray}
where $g_V^i$ and $g_A^i$ are the vector and axial vector couplings of respective fermions in SM.

\par If the decay and annihilation rates are much smaller than
the Hubble expansion rate at the temperature of the order of
Elctro-Weak (EW) symmetry-breaking scale, the spin-$3/2$ DM
particle $\chi$ will never be in thermal equilibrium. The 
Hubble expansion rate at a temperature $T$ is given by $H(t) \approx 1.66 g^* T^2/m_{\rm Pl}$ where $g^\star $ is the effective number of relativistic degrees of freedom at
the temperature $T$. The decay $Z \to \chi \overline\nu_e+\overline\chi\nu_e$ comes into play only 
below EW symmetry breaking phase transition temperature 
$T_{EW} \sim 150~ \rm{GeV}$. The $Z$ bosons go out of the equilibrium below 
roughly 5 GeV, the other $SM$ fermions remain in equilibrium much below this temperature. 
\par The decay and annihilation rates can be estimated from Eqs. \eqref{ZDecay}-\eqref{vectorFusion}. The $Z$ decay rate is given by
\begin{equation}
\Gamma(Z \to \chi \overline\nu_e + \overline\chi \nu_e)
\approx 7.7 \times 10^{21}~ C^2 \left[\frac{\Lambda}{1\, {\rm GeV}}\right]^{-4} ~ {\rm GeV}.
\end{equation}
The leading terms in the cross-section corresponding to $\chi$ production
through $f\overline f$ annihilation and $W $fusion processes are given by
\begin{equation}
\sum_i \sigma(f_i \overline f_i  \to \chi \overline\nu_e + \overline\chi \nu_e)
\approx  2.3 \times 10^{13}~ C^2\, \left[\frac{s}{\rm 1\, GeV^2}\right]\,\left[\frac{\Lambda}{1\, {\rm GeV}}\right]^{-4}\,\, {\rm GeV}^{-2}
\end{equation}
and
\begin{equation}
\sigma (W^+ W^-  \to \chi \overline\nu_e + \overline\chi \nu_e)
\approx  6.96 \times 10^{3}~  C^2\,\, \left[\frac{s}{\rm 1\, GeV^2}\right]^3\, \left[\frac{\Lambda}{1\, {\rm GeV}}\right]^{-4}\,\, {\rm GeV}^{-2},
\end{equation}
respectively. One can obtain the constraint on the effective coupling $C/\Lambda^2$ by demanding  the thermal average $\Gamma ( Z\to  \chi \overline\nu_e + \overline\chi \nu_e)$, $ \left\langle \sigma\left( \sum _ff \overline f\to  \chi \overline\nu_e + \overline\chi \nu_e\right)|
n v\right\rangle$ and $\left\langle \sigma( W^+ W^-\to  \chi \overline\nu_e + \overline\chi \nu_e)|n v\right\rangle$  to be  less than the $H(T)$ for $T \sim 150$ GeV ({\it i.e.} at the EW phase transition temperature). Therefore,  using $g^\star =113.45$, we obtain
\begin{equation}
C\,\left[ \frac{\Lambda}{\rm 1\, GeV}\right]^{-2} \le  0.9 \times 10^{-17} \rm{~ from~
 Z~ decay,}
\end{equation}
\begin{equation}
C\,\left[  \frac{\Lambda}{1\,{\rm GeV}}\right]^{-2} \le 1.0 \times 10^{-19} {\rm ~  from~}\,
  \left( \Sigma_f f \overline f \to \chi\overline\nu_e+\overline\chi\nu_e\right),
\end{equation}
and
\begin{equation}
C\,\left[\frac{\Lambda}{\rm 1\, GeV}\right]^{-2} \le 2.4 \times 10^{-19} {\rm ~ from~}\,
W^+W^-\to \chi\overline\nu_e+\overline\chi\nu_e.
\end{equation}
The thermal averaged cross-sections $\left\langle \sigma \vert n v \right\rangle $
are estimated using the relation $s=4 \left\langle E\right\rangle^2 $ where
$\left\langle E\right\rangle =3.15 ~\rm{T} ~\rm{and}~ 2.7~\rm{T}$, and 
$n=\displaystyle\frac{3}{4}\frac{\zeta(3)}{\pi^2}gT^3$ and 
$\displaystyle\frac{\zeta(3)}{\pi^2}gT^3$ for Fermi-Dirac and Bose-Einstein 
particles, respectively.

\par The relic density of the spin-$3/2$ DM $\chi$ can be evaluated by 
solving the Boltzmann equation for the evolution of the number density
$n_\chi$ of DM $\chi$ and is given by
\begin{eqnarray}
\dot{n_{\chi}} + 3 H n_{\chi} &=&-\langle\Gamma (Z\to \chi\overline\nu_e)\rangle (n_{\chi}^2-n_0^2)-\Sigma_f \left\langle \sigma \left(\sum_f f \overline f\to \chi\overline\nu_e\right)|v\right\rangle (n_{\chi}n_{\nu_e}^0-{n^f_0}^2) \nonumber\\&&
 -\left\langle \sigma (W^+ W^-\to \chi\overline\nu_e)|v\right\rangle (n_{\chi}n_{\nu_e}^0-{n^W_0}^2).
\end{eqnarray}
Here, $n^i_0$ is the equilibrium number density of species $i$. The region of
validity of the equation is when all the SM particles are in thermal equilibrium unlike the DM candidate
$\chi$  which is realized for  $T \lesssim (15-20)\, m_i$.
We can than put $n_{\chi} =0$ in the R.H.S. of this equation. 
Changing the variable from time to temperature, the equation can be put in the form:
\begin{equation}
\frac{df_{\chi}}{dz}=\frac{\Gamma(Z \to \chi \overline{\nu}_e)}{K m_Z^2}
z f_0^Z +
\sum_i \frac{\langle \sigma (f_i \overline{f_i}\to \chi \overline{\nu}_e)|v\rangle}{K Z^2} (f_0^i)^2,
\end{equation}
where $z=m_Z/T$, $f_\chi = n_\chi/T^3$, $f_0^i = n_0^i/ T^3$, and $K= 1.66 \sqrt{g^\star}/m_{Pl}$. We use Boltzmann distribution functions for both the fermions and bosons,
\textit{i.e.} 
\begin{equation}
f^i_0 (m_i/T) =f^i_0 \left(\frac{m_i}{m_Z}\frac{m_Z}{T}\right)= f^i_0 (x_i z) 
=\frac{g_i}{2 \pi^2} \int_0^{\infty} p^2 e^{-\sqrt{p^2+x^2_iz^2}}dp . 
\end{equation}
The thermal averaged decay rate and annihilation cross-sections can be expressed, following Ref. \cite{Gondolo:1990dk,book}, as 
\begin{equation}
\langle \Gamma (Z \to \chi \overline{\nu_e}) \rangle =  \Gamma (Z \to \chi \overline{\nu}_e)\frac{K_1(z)}{K_2 (z)}\label{avGamma}
\end{equation} 

and
\begin{equation}
\langle \sigma(f_i \overline{f_i}\to \chi \overline{\nu}_e) \rangle =
\displaystyle
\frac{1}{8 m_i^4 T  K^2_2 \left(\frac{m_i}{T}\right)} \int_{4 m_i^2}^{\infty}
\sigma(f_i \overline{f_i}\to \chi \overline{\nu})(s-4 m_i^2) \sqrt{s} 
K_1 \left( \frac{\sqrt{s}}{T}\right)\,\, ds.\label{avSigma}
\end{equation} 
Here, $K_{1,2}(x)$ are the modified Bessel functions. In terms of scaled number
density defined as $N_{\chi} = f_{\chi}Km_Z$ and by using the expressions
for thermal averaged decay width and the annihilation cross-sections given
in Eqs. (\ref{avGamma}) and (\ref{avSigma}), the Boltzmann equation can be
written as
\begin{eqnarray}
\frac{dN_{\chi}}{dz} & = & \Gamma (Z \to \chi \overline{\nu}_e) 
\frac{K_1(z)}{K_2(z)}\frac{1}{m_Z} z f_0^z (z) \nonumber \\
&+& \sum_i \frac{1}{8} \frac{z}{x_i^4} \frac{m_Z^2}{K^2_2(x_i z)}
\int_{4 x_i^2}^{\infty} (y-4 x_i^2) \sqrt{y} K_1 (\sqrt{y}z) 
\sigma (f_i \overline{f_i} \to \chi \overline{\nu}_e) \frac{1}{z^2} 
\left( f_0^i (x_i z) \right)^2\,\, dy.\nonumber\\
\label{Boltzmann}
\end{eqnarray}
We solve the above Boltzmann equation for the scaled number density $N_{\chi}$
of spin-$3/2$ dark matter particle $\chi$ for $0.6<z<18$ corresponding to
$\text{5 GeV < T < 150 GeV}$.

\par The contribution of the proposed $7.1$ keV spin-$3/2$ fermion $\chi$ to the relic
dark matter density is obtained by numerically solving the Boltzmann Eq. \eqref{Boltzmann}
from the electroweak phase transition temperature to the freeze-out temperature
of $W's$ and $Z's$. The scaled number density $N_{\chi}$'s for the leading processes that maintain the
dark matter $\chi$ in equilibrium with the rest of SM plasma are obtained to be
\begin{eqnarray}
N_{\chi} \left(\Gamma (Z) \right) \simeq C^2\times 10^{19}\left[\frac{\Lambda}{1\, {\rm GeV}}\right]^{-4};\,\,\,
N_{\chi} \left(\sum_f\sigma (f\overline{f}) \right) \simeq C^2\times 10^{20}\left[\frac{\Lambda}{1\, {\rm GeV}}\right]^{-4};\nonumber\\
{\rm and}\,\,N_{\chi} \left(\sigma (W^+W^-) \right) \simeq C^2\times 10^{23}\left[\frac{\Lambda}{1\, {\rm GeV}}\right]^{-4};
\end{eqnarray}
for the $Z$-decay, fermion-antifermion annihilation and $W^\pm$  fusion processes,
respectively. 

We find that the contribution of spin-$3/2$ DM fermion to the relic density from $W^\pm$ boson
fusion process is about three order of magnitudes greater than the contribution
from the rest of the processes. For our estimate of the dark matter density, we use the
$N_{\chi} \approx C^2\times  10^{23}\, \left[\Lambda/\,1\, {\rm GeV}\right]^{-4}$.  Thus, the number density
of $\chi$ at the electroweak phase transition temperature is given by
$\left. n_\chi/\, T^3\right\vert_{T=T_{EW}} \approx N_{\chi}/\, (K\, M_z) 
\sim 10^{23}/\,(K \,M_z)$. The number density of $\chi$'s as the Universe cools to the present day is estimated to be $\left.n_{\chi} \right\vert_{T_0} \sim T_0^3\times 10^{23}/ (\zeta \,K\, m_Z)$
where $T_0$ is the present day temperature ($T_0 =2.73 K$) and 
$\zeta = g^*(T_{EW})/ \,g^*(T_0) \sim 33.85$. 
The present day dark matter relic density $\left.\rho_\chi \right\vert_{T_0} \approx C^2\,\times 10^{-5} \,\left[\Lambda/\,1\, {\rm GeV}\right]^{-4}\,\, {\rm GeV}^4
$ is then 
obtained by multiplying the number density $n_{\chi}$ with its mass $m_\chi$. Since, the critical dark matter density ${\rho_\chi}_c \sim 8.1~ h^2~ 10^{-47}~ {\rm GeV}^{4}$, the $\Omega_{\chi} = \rho_{\chi}/{\rho_\chi}_c$ is computed as 
\begin{equation}
\Omega_{\chi} h^2 \approx 0.11 \times 10^{42}\,\,\, C^2\,\,\, \left[\frac{\Lambda}{1\, {\rm GeV}}\right]^{-4}.
\end{equation}
 However, the desired value of $\Omega_{\chi} h^2 \sim \Omega_{DM} =0.11$ will be obtained for
$C \left[\Lambda/\,1\, {\rm GeV}\right]^{-2}\approx 10^{-21}$.

\section{Supernova Energy Loss}
\label{stellarloss}
The $7.1$ keV spin-$3/2$ dark matter $\chi$ can be a source of 
significant energy loss in the supernova core. The emission rates for SN 1987 A 
have been extensively studied for weakly interacting DM candidate particles like 
axions, gravitinos, right handed neutrinos, majorons, low mass neutralinos, Goldstone bosons etc. 
in new physics models. Constraints have been put on the properties  and 
interactions of these particles \cite{Kolb:1988pe, Raffelt:1987yt,   Cullen:1999hc, Barger:1999jf, Dutta:2007tz, Keung:2013mfa, Barbieri:1988av}. The SN bound on neutrino magnetic 
dipole moment have been one of the tightest \cite{Goyal:1995yd}. In our estimate of the 
constraints on the parameters of our model, we would use the Raffelt 
criterion \cite{raffeltbook} that new source of cooling should not exceed the emissivity $\dot 
\epsilon /\rho=10^{19}$ ergs per gm per sec. The main source of $\chi$ pair 
production in the core of SN is through the $\chi\,\overline\nu_e$ and/ or $  \overline\chi\,\nu_e$ production   processes.
The emissivity {\it i.e.} the energy emitted per unit time and volume, is
\begin{equation}
\dot{\epsilon} = \int \prod_{i=1}^{4}\,\frac{d^3p_i}{(2\pi)^3\,2 \,E_i}\, (2\pi)^4 \,\delta^4\left(p_{e^+}+p_{e^-}-p_{\overline\nu_e}-p_\chi\right)f_{1}\,f_{2}\,\left(1-f_{3}\right)\,\left(1-f_4\right)E_{\chi} \overline{\left\vert M\right\vert^2},
\end{equation} 
where $\overline{\left\vert M\right\vert^2}$ is the matrix element squared, summed over the initial and
final states and $ f_i\equiv \left[\exp(E_i-\mu_i)/T+1\right]^{-1}$ is the 
Fermi-Dirac distribution for the $i$th particle.

In the supernova core immediately after the collapse, the temperature is 
high being of the order of tens of MeV. Even though the nucleons are 
nearly non-degenerate, the electrons are degenerate and the neutrinos are
trapped. The core has a fixed value of the lepton number. Thus, there also 
exists a sub-dominant energy loss process via the neutrino pair annihilation 
$\nu\overline{\nu} \to \chi\overline{\nu_e}+\overline{\chi}\nu_e$.
Since, the coupling of the dark matter particle $\chi$ to SM fermions is
extremely weak, the $\chi$'s once produced freely stream out of the SN core, 
their mean free path being greater than the core radius. We thus have 
$\mu_{\chi} \approx 0$. Carrying out the phase space integrals and making a 
change in variables from $E_1$, $E_2$, $\theta$ to $E_+=E_1+E_2$, 
$E_-=E_1-E_2$ and $s = 2 m_e^2 + 2 E_1 E_2 - 2 \vec{p_1}.\vec{p_2}\, \cos \theta$,
we get
\begin{equation}
\dot{\epsilon} = \frac{1}{2 \pi^3} \int_{4 m_e^2}^{\infty}ds 
\int_{\sqrt{s}}^{\infty} dE_+ 
\int_{-\sqrt{E^2+s}}^{\sqrt{E^2+s}}dE_-\,\,
s \frac{E_+ +  E_-}{2} f_1 f_2 \,\sigma (e^+e^- \to \chi \overline{\nu}_e).
\label{emissexp}
\end{equation}

In deriving the above expression, we have neglected the Pauli blocking terms
for the final state particles $\chi$ and $\nu_e$ which is an excellent approximation
for $\overline{\nu}_e$, $\chi$ and $\overline{\chi}$.  We have similar expression for the
process $\nu_e\overline{\nu}_e \to \chi \overline{\nu}_e + \overline{\chi}\nu_e$. The
cross-section for these processes has been evaluated in Eq. \eqref{fermionAnnihilation}.

The core density lies anywhere between $3 \times 10^{13}$ to 
$3 \times 10^{14} \rm{gm/cc}$. At a core temperature of about $40~\rm{MeV}$,
the electron chemical potential is $\mu_e \approx 200 ~\rm{MeV}$ and 
$\mu_e -\mu_{\nu_e} \approx 50 ~\rm{MeV}$. In our estimate of the 
energy loss, we consider the core density to be $3 \times 10^{14}~\rm{GeV}$ with
a core temperatures $30 ~(50) ~ \rm{MeV}$ and electron and neutrino chemical 
potentials 200 (150) MeV and 150 (100) MeV, respectively, and evaluate the
energy loss integral numerically.
\par Constraints from supernova cooling are obtained by numerical integration of the
emissivity expression \eqref{emissexp} for the process $e^+e^- \to \chi \overline{\nu}_e+ \overline{\chi} \nu_e$ at $T = 30~\rm{MeV}$ and electron chemical potential
$\mu_e = 200 ~\rm{MeV}$, we obtain
\begin{equation}
\frac{\dot{\epsilon}(e^+e^- \to \chi \overline{\nu}_e
+ \overline{\chi} \nu_e)}{\rho_{\rm core}} = 2.2 \times 10^{53}\,\, C^2\,\,\left[\frac{\Lambda}{1\, {\rm GeV}}\right]^{-4} \,\, {\rm ergs/gm/s}
\end{equation} 
where we have taken the core density $\rho_{\rm core} $ to be about $3 \times 10^{14} \rm{gm/cc}$.
Requirement of $\frac{\dot{\epsilon}}{\rho_{\rm core}} < 10^{19} \rm{ergs/gm/cc}$ constrains
\begin{equation}
C\,\, \left[\frac{\Lambda}{1\, {\rm GeV}}\right]^{-2}~  \le 6.7 \times 10^{-18}.
\end{equation}
Core temperature of $50~\rm{MeV}$ and $\mu_e =250 MeV$ results in a somewhat
tighter constraint $C \, \left[\Lambda/\,1\, {\rm GeV}\right]^{-2} \le 10^{-18}$. The contribution 
from the process $\nu_e \overline{\nu}_e \to \chi \overline{\nu}_e +\overline{\chi} \nu_e$ is
totally negligible being roughly 10 orders of magnitude smaller compared to the
annihilation process.
\section{Results and Discussion}
\label{results}
\subsection{Summary}
\begin{figure}
 \centering
\vskip -14cm
\includegraphics[width=\textwidth,clip]{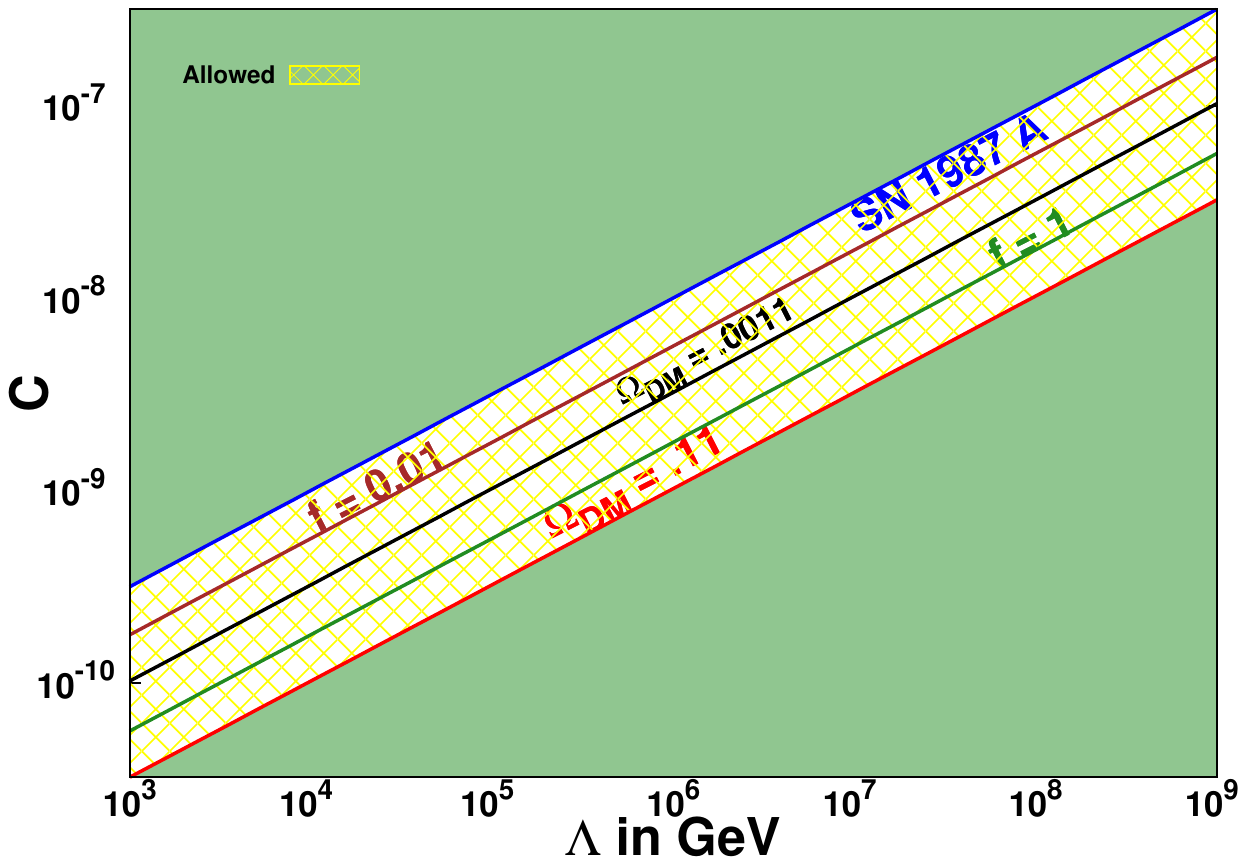}
\vskip -1 cm
\caption{\small \em{Combined  constraints on the coupling $C $ and cut-off scale $\Lambda $ from contribution to the relic density as WDM, the rapid cooling of supernova SN 1987 A and decay of the spin-$3/2$ particle. Allowed region (with yellow lines) of the parameter space is bounded by the maximum DM relic density constraint $\Omega_{DM}=0.11$  and from supernova cooling of SN 1987 A. The region in green is forbidden.}}
\label{fig:constraints}
\end{figure}

We  summarize the constraints on the parameters of our $7.1$ keV 
spin-$3/2$ dark matter particle $\chi$ from cosmological and astrophysical observations.
 We observe that the ratio of the coupling and the square of the cut-off scale $C/\Lambda^2$ associated with spin-$3/2$ particle  $\chi$  of mass  $7.1$ keV  can be constrained as
\begin{itemize}
\item $C\, \left[\frac{\Lambda}{1\, {\rm GeV}}\right]^{-2} \lesssim  10^{-21}$  from the consideration of $\chi$ as a WDM candidate accounting for the entire observed  relic dark matter density $\Omega _\chi h^2 = \Omega_{DM}= 0.11$, 
\item  $ C\,\,\left[\frac{\Lambda}{1\, {\rm GeV}}\right]^{-2}\lesssim 6.7 \times 10^{-18}
$ from the the rapid cooling of the supernova through the emission of $\chi$ and,
\item  $ C\,\, \left[\frac{\Lambda}{1\, {\rm GeV}}\right]^{-2} \approx 2.4 \times 10^{-20}$ from the lifetime of $\chi$ through its decay $\chi\to \nu_e \gamma$. 
\end{itemize}
These combined constraints  on the parameter space of coupling $C$ and the cut-off  scale $\Lambda$ arising from its appropriate lifetime,  contribution to relic density and supernova cooling are shown in figure \ref{fig:constraints}. The curves marked $f=0.01$ and $f=1$ correspond to the life time  $\tau_\chi$ required for the observed X-ray flux for WDM $\chi$ contribution $\Omega_\chi h^2=0.01 \times \Omega_{DM}$ and $\Omega_\chi h^2=\Omega _{DM}$  respectively. We find that the constraints from the cooling of supernova 1987A  and the DM relic density $\Omega_\chi h^2=0.11$ enclose an allowed band (shaded with yellow lines in the figure) in the parameter space spanned by  $C$ and  $\Lambda$. The parameter space  shaded in green is forbidden.

\par We thus see that a minimal extension of the SM by adding a spin-$3/2$
SM singlet with mass $7.1$ keV can account for the dark matter in 
the Universe, while at the same time explaining the $3.55$ keV X-ray
line in the galactic X-ray spectrum through its decay 
$\chi \to \nu \gamma$.
\subsection{Outlook}
\label{outlook}
Recently, superconducting detectors are proposed for direct detection of light
DM particles of mass as low as $1$ keV through electron recoil from DM-electron
scattering in superconductors \cite{Hochberg:2015pha}. It will be worthwhile to study the
DM model discussed in this article to compute the DM scattering rates with
electrons in a superconducting environment where electrons are highly 
degenerate and the scattering is inhibited by the Pauli blocking and to 
explore the feasibility of detecting the proposed DM particle. We leave this
for the future work.\\

\acknowledgments
 We would like to thank the referee for drawing our attention to a recent 
 analysis of the X-ray spectrum lines observed by XMM-Newton \cite{Jeltema:2014qfa} and for his constructive suggestions to improve the manuscript.
SD would like to thank IUCAA, Pune for hospitality where part of this work was 
completed. AG would like to acknowledge CSIR (ES) Award for partial support. 
SK acknowledges financial support from the DST project FTP/PS-123/2011.


\begin{thebibliography}{99}
\bibitem{Bulbul:2014sua}
  E.~Bulbul, M.~Markevitch, A.~Foster, R.~K.~Smith, M.~Loewenstein and S.~W.~Randall,
  Astrophys.\ J.\  {\bf 789}, 13 (2014).

\bibitem{Boyarsky:2014jta} 
  A.~Boyarsky, O.~Ruchayskiy, D.~Iakubovskyi and J.~Franse,
  Phys.\ Rev.\ Lett.\  {\bf 113}, 251301 (2014).
\bibitem{Ishida:2014dlp} 
  H.~Ishida, K.~S.~Jeong and F.~Takahashi,
  Phys.\ Lett.\ B {\bf 732}, 196 (2014).

\bibitem{Chakraborty:2014tma} 
  S.~Chakraborty, D.~K.~Ghosh and S.~Roy,
  JHEP {\bf 1410}, 146 (2014).
\bibitem{Modak:2014vva} 
  K.~P.~Modak,
  JHEP {\bf 1503}, 064 (2015).

\bibitem{Abazajian:2014gza} 
  K.~N.~Abazajian,
  Phys.\ Rev.\ Lett.\  {\bf 112}, no. 16, 161303 (2014).

 \bibitem{Biswas:2015sva} 
  A.~Biswas, D.~Majumdar and P.~Roy,
  JHEP {\bf 1504}, 065 (2015).

\bibitem{Liew:2014gia} 
  S.~P.~Liew,
  JCAP {\bf 1405}, 044 (2014).
  [arXiv:1403.6621 [hep-ph], arXiv:1403.6621].
\bibitem{Kolda:2014ppa} 
  C.~Kolda and J.~Unwin,
  Phys.\ Rev.\ D {\bf 90}, 023535 (2014).

\bibitem{Higaki:2014zua} 
  T.~Higaki, K.~S.~Jeong and F.~Takahashi,
  Phys.\ Lett.\ B {\bf 733}, 25 (2014).

\bibitem{Kong:2014gea} 
  J.~C.~Park, S.~C.~Park and K.~Kong,
  Phys.\ Lett.\ B {\bf 733}, 217 (2014).

\bibitem{Covi:2009pq} 
  L.~Covi and J.~E.~Kim,
  New J.\ Phys.\  {\bf 11}, 105003 (2009).


\bibitem{Lee:2014xua} 
  H.~M.~Lee, S.~C.~Park and W.~I.~Park,
  Eur.\ Phys.\ J.\ C {\bf 74}, 3062 (2014).

\bibitem{Krall:2014dba} 
  R.~Krall, M.~Reece and T.~Roxlo,
  JCAP {\bf 1409}, 007 (2014).

\bibitem{Choi:2014tva} 
  K.~Y.~Choi and O.~Seto,
  Phys.\ Lett.\ B {\bf 735}, 92 (2014).

\bibitem{Babu:2014pxa} 
  K.~S.~Babu and R.~N.~Mohapatra,
  Phys.\ Rev.\ D {\bf 89}, 115011 (2014).

\bibitem{Burges:1983zg} 
  C.~J.~C.~Burges and H.~J.~Schnitzer,
  Nucl.\ Phys.\ B {\bf 228}, 464 (1983);
  J.~H.~Kuhn and P.~M.~Zerwas,
  Phys.\ Lett.\ B {\bf 147}, 189 (1984).


\bibitem{ArkaniHamed:1998rs} 
  N.~Arkani-Hamed, S.~Dimopoulos and G.~R.~Dvali, Phys.\ Lett.\ B {\bf 429}, 263 (1998);  N.~Arkani-Hamed, S.~Dimopoulos and J.~March-Russell
  Phys.\ Rev.\ D {\bf 63}, 064020 (2001);  I.~Antoniadis, N.~Arkani-Hamed, S.~Dimopoulos and G.~R.~Dvali, Phys.\ Lett.\ B {\bf 436}, 257 (1998).


\bibitem{Randall:1999ee} 
  L.~Randall and R.~Sundrum,
  Phys.\ Rev.\ Lett.\  {\bf 83}, 3370 (1999).

\bibitem{Lemoine:2005hu} 
  M.~Lemoine, G.~Moultaka and K.~Jedamzik,
  Phys.\ Lett.\ B {\bf 645}, 222 (2007).
\bibitem{Jedamzik:2005ir} 
  K.~Jedamzik, M.~Lemoine and G.~Moultaka,
  Phys.\ Rev.\ D {\bf 73}, 043514 (2006).
\bibitem{Baltz:2001rq} 
  E.~A.~Baltz and H.~Murayama,
  JHEP {\bf 0305}, 067 (2003).
\bibitem{Fujii:2002fv} 
  M.~Fujii and T.~Yanagida,
  Phys.\ Lett.\ B {\bf 549}, 273 (2002).
\bibitem{Moultaka:2006su} 
  G.~Moultaka,
  Acta Phys.\ Polon.\ B {\bf 38}, 645 (2007).
\bibitem{Gorbunov:2008ui} 
  D.~Gorbunov, A.~Khmelnitsky and V.~Rubakov,
  JHEP {\bf 0812}, 055 (2008)

\bibitem{Ding:2012sm} 
  R.~Ding and Y.~Liao,
  JHEP {\bf 1204}, 054 (2012).
\bibitem{Ding:2013nvx} 
  R.~Ding, Y.~Liao, J.~Y.~Liu and K.~Wang,
  JCAP {\bf 1305}, 028 (2013).

\bibitem{Savvidy:2012qa} 
  K.~G.~Savvidy and J.~D.~Vergados,
  Phys.\ Rev.\ D {\bf 87}, no. 7, 075013 (2013).

\bibitem{Jeltema:2014qfa} 
  T.~E.~Jeltema and S.~Profumo,
  Mon.\ Not.\ Roy.\ Astron.\ Soc.\  {\bf 450}, no. 2, 2143 (2015).
\bibitem{Maccio':2012uh} 
  A.~V.~Maccio, O.~Ruchayskiy, A.~Boyarsky and J.~C.~Munoz-Cuartas,
  Mon.\ Not.\ Roy.\ Astron.\ Soc.\  {\bf 428}, 882 (2013).
\bibitem{Anderhalden:2012qt} 
  D.~Anderhalden, J.~Diemand, G.~Bertone, A.~V.~Maccio and A.~Schneider,
  JCAP {\bf 1210}, 047 (2012).
\bibitem{Abazajian:2001vt} 
  K.~Abazajian, G.~M.~Fuller and W.~H.~Tucker,
  Astrophys.\ J.\  {\bf 562}, 593 (2001).
\bibitem{Lola:2007rw} 
  S.~Lola, P.~Osland and A.~R.~Raklev,
  Phys.\ Lett.\ B {\bf 656}, 83 (2007)
\bibitem{Bomark:2014yja} 
  N.-E.~Bomark and L.~Roszkowski,
  Phys.\ Rev.\ D {\bf 90}, 011701 (2014).

\bibitem{Cowsik:1972gh} 
  R.~Cowsik and J.~McClelland,
  Phys.\ Rev.\ Lett.\  {\bf 29}, 669 (1972).

\bibitem{Gondolo:1990dk} 
  P.~Gondolo and G.~Gelmini,
  Nucl.\ Phys.\ B {\bf 360}, 145 (1991).


\bibitem{book} E.~W.~Kolb and M.~S.~Turner, {\it The Early Universe}, Addison-Wesley publishing Company, (1990).
\bibitem{Kolb:1988pe} 
  E.~W.~Kolb and M.~S.~Turner,
  Phys.\ Rev.\ Lett.\  {\bf 62}, 509 (1989);   
  E.~W.~Kolb and M.~S.~Turner,
  Phys.\ Rev.\ D {\bf 36}, 2895 (1987).


\bibitem{Raffelt:1987yt} 
  G. G. ~Raffelt and D.~Seckel,
  Phys.\ Rev.\ Lett.\  {\bf 60}, 1793 (1988).

\bibitem{Cullen:1999hc} 
  S.~Cullen and M.~Perelstein,
  Phys.\ Rev.\ Lett.\  {\bf 83}, 268 (1999).
\bibitem{Barger:1999jf} 
  V.~D.~Barger, T.~Han, C.~Kao and R.~J.~Zhang,
  Phys.\ Lett.\ B {\bf 461}, 34 (1999).

\bibitem{Dutta:2007tz} 
  S.~Dutta and A.~Goyal,
  JCAP {\bf 0803}, 027 (2008).

\bibitem{Keung:2013mfa} 
  W.~Y.~Keung, K.~W.~Ng, H.~Tu and T.~C.~Yuan,
  Phys.\ Rev.\ D {\bf 90}, no. 7, 075014 (2014).


\bibitem{Barbieri:1988av} 
  R.~Barbieri and R.~N.~Mohapatra,
  Phys.\ Rev.\ D {\bf 39}, 1229 (1989).


\bibitem{Goyal:1995yd} 
  A.~Goyal, S.~Dutta and S.~R.~Choudhury,
  Phys.\ Lett.\ B {\bf 346}, 312 (1995).

\bibitem{raffeltbook} G. G. Raffelt, {\it Stars as Laboratories
for Fundamental Physics}. University of Chicago Press, Chicago (1996).

\bibitem{Hochberg:2015pha} 
  Y.~Hochberg, Y.~Zhao and K.~M.~Zurek,
  arXiv:1504.07237 [hep-ph].

\end{thebibliography}
\end{document}